\documentclass[twocolumn]{aastex62}
\usepackage{multirow}
\usepackage{rotating}
\usepackage{xfrac}

\newcommand\Msun{M$_{\odot}$}
\newcommand\flux{erg s$^{-1}$ cm$^{-2}$}
\newcommand\lum{erg s$^{-1}$}

\graphicspath{{./}{figures/}}

\accepted{14 March 2019}
\submitjournal{AAS Journals}

\shorttitle{X-rays from Stellar Bow Shocks}
\shortauthors{Binder et al.}

\begin{document}

\title{Searching for Faint X-ray Emission from Galactic Stellar Wind Bow Shocks}

\correspondingauthor{Breanna A. Binder}
\email{babinder@cpp.edu}

\author[0000-0002-4955-0471]{Breanna A. Binder}
\affil{Department of Physics \& Astronomy, California State Polytechnic University, 3801 W. Temple Ave., Pomona, CA 91768, USA}

\author{Patrick Behr}
\affil{Department of Physics \& Astronomy, California State Polytechnic University, 3801 W. Temple Ave., Pomona, CA 91768, USA}

\author{Matthew S. Povich}
\affil{Department of Physics \& Astronomy, California State Polytechnic University, 3801 W. Temple Ave., Pomona, CA 91768, USA}

\begin{abstract}

We present a stacking analysis of 2.61 Msec of archival {\em Chandra} observations of stellar wind bow shocks. We place an upper limit on the X-ray luminosity of IR-detected bow shocks of $<2\times10^{29}$ \lum, a more stringent constraint than has been found in previous archival studies and dedicated observing campaigns of nearby bow shocks. We compare the X-ray luminosities and $L_X/L_{\rm bol}$ ratios of bow shock driving stars to those of other OB stars within the {\em Chandra} field of view. Driving stars are, on average, of later spectral type than the ``field of view'' OB stars, and we do not observe any unambiguously high $L_X/L_{\rm bol}$ ratios indicative of magnetic stars in our sample. We additionally asses the feasibility of detecting X-rays from stellar wind bow shocks with the proposed {\em Lynx} X-ray Observatory. If the X-ray flux originating from the bow shocks is just below our {\em Chandra} detection limit, the nearest bow shock in our sample (at $\sim$0.4 kpc with an absorbing column of $\sim10^{21}$ cm$^{-2}$) should be observable with {\em Lynx} in exposure times on the order of $\sim$100 kiloseconds. 
\end{abstract}

\keywords{Stars: massive --- Stars: winds, outflows --- circumstellar matter --- X-rays: ISM}

\section{Introduction} \label{sec:intro}
Stellar wind bow shocks produced by runaway OB stars ($M>$ 8 \Msun) are believed to be a major source of high-energy emission in the Milky Way \citep{Mohamed+12,delValle+15,Toala+16,delValle+18}. Non-thermal emission arises from relativistic particles (mainly electrons) being accelerated by a magnetic field at the shock front \citep{delValle+12,Meyer+17} via first-order Fermi acceleration; models of this process typically use the magnetic field strength of the ambient ISM \citep[$\sim$few $\mu$G;][]{Meyer+17} or equipartition arguments with respect to the kinetic power from the driving star's stellar wind \citep[$\sim$tens of $\mu$G;][]{delValle+12}. The electrons then participate in inverse Compton scattering with IR photons emitted by swept-up dust in the bow shock, scattering the IR photons up to X-ray and gamma ray energies \citep{Peri+12, delValle+18}.

These models, however, predict only faint X-ray luminosities, $\sim(1-10)\times10^{29}$ erg s$^{-1}$ \citep[e.g., ][]{Pereira+16, Meyer+17, DeBecker+17, delValle+18}, making direct X-ray observations of bow shocks difficult. Although \citet{LopezSantiago+12} reported the first high energy detection of the AE Aurigae bow shock, it was later shown by \citet{Toala+17} that the X-ray emission was not spatially coincident with the IR bow shock. \citet{Toala+16} report a detection of diffuse, thermal X-ray emission surrounding $\zeta$ Oph using {\em Chandra}, although pile-up during the observation makes direct interpretation of the X-ray spectrum difficult. Non-thermal radio emission has been observed in BD +43$^{\circ}$3654 \citep{Benaglia+10}, although observations with {\em Suzaku} and {\em XMM-Newton} have only yielded upper limits on the X-ray luminosity \citep[3$\sigma$ upper limit of 1.1$\times$10$^{32}$ erg s$^{-1}$;][]{Terada+12, Toala+16}. Recent 2D hydrodynamical models by \citet{delValle+18} suggest that only a very small fraction of the total stellar wind kinetic power ($\sim 10^{-5}$) is converted into nonthermal emission, and that such emission is more likely to be detected at radio wavelengths. At these faint luminosities, any hope of directly observing X-rays from bow shocks $>$0.5 kpc from the Sun requires prohibitively long exposure times.

In contrast, the massive stars driving the bow shocks may appear as X-ray point sources that are detectable in modest exposure times. Roughly 10\% of OB stars exhibit strong (often dipolar) magnetic fields \citep{Petit+13,udDoula+16}. In these systems, charged particles from the stellar wind become trapped and channeled along magnetic field lines, leading to magnetically confined wind shocks \citep{Babel+97a,Babel+97b}. These stars have systematically higher X-ray luminosities and $L_{\rm X}/L_{\rm bol}$ ratios than non-magnetic OB stars. The prototypical example is $\theta^1$ Ori C \citep{Gagne+05}, and magnetic fields have also been detected in Tr~16-22 and Tr~16-13 in the Carina Nebula \citep{Naze+12}. Stellar wind bow shock nebulae formation, however, is likely not affected by the magnetic surface properties of driving stars, as the radial dipolar component of the stellar $B$-field falls off as $\sim1/r^2$. The stellar $B$-field is thus negligible at the standoff radius $R_0$ (typically at $\sim0.1-1$ pc).

We have leveraged 91 archival {\em Chandra} X-ray observations (with a total exposure time of 2.61 Msec) containing 60 IR-bright Galactic stellar-wind bow shocks. The majority ($\sim$70\%) of these bow shocks are located in apparently isolated environments, making them good candidate runaway stars. By stacking the X-ray images, we have created the deepest X-ray exposure of IR stellar wind bow shocks to date. We additionally performed a comparison of the X-ray properties of bow shock driving stars to other massive stars detected within the {\em Chandra} field of view. In Section~\ref{section:observations} we describe the IR bow shock sample, the {\em Chandra} archival observations we use, and describe our data reduction and processing procedures (including X-ray point source detection). In Section~\ref{section:bow_shocks_stacking} we describe our bow shock stacking analysis, and Section~\ref{section:stars} presents a comparison of the X-ray properties of driving stars to other OB stars. We discuss our results in Section~\ref{section:discussion} and prospects for observations with future X-ray facilities, and summarize our findings in Section~\ref{section:conclusions}.

\section{Sample Selection}\label{section:observations}
The largest published catalog of mid-IR stellar bow shocks contains 709 sources identified in {\em Spitzer} and {\em WISE} images of the Galactic Plane \citep{Kobulnicky+16}, and the latest data release from the citizen science initiative the Milky Way Project (MWP; Jayasinghe et al. \textit{in prep}) has found an additional 282 IR stellar wind bow shocks that we include in our initial Galactic bow shocks sample.

The massive OB driving stars of these stellar wind bow shocks may also appear as X-ray point sources. The superior angular resolution of the {\em Chandra} X-ray Observatory is therefore necessary to separate potential X-ray emission from the bow shock driving star from the bow shock itself. We searched the {\em Chandra} archive for ACIS observations containing the position of at least one IR-identified bow shock within 5$^{\prime}$ of the nominal aim point. This radius was chosen to ensure reasonably small (${\la}2\arcsec$) and symmetric PSFs at the position of the bow shock. The search returned 91 archival observations (containing 60 unique bow shocks) with a total of 2.61 Msec exposure time. In Table~\ref{table:observation_log}, we summarize the observation identification numbers (hereafter referred to as ``ObsIDs''), the date of the observation, which ACIS instrument was used (S or I), the nominal aim-point, the effective exposure time (see next section), and the number of bow shocks contained within 5$^{\prime}$ of the nominal aim point. Most archival observations were originally used to study massive stars, star clusters, and young stellar objects in H II regions, such as the Carina Nebula and M17. Other observations targeted pulsars and their resulting wind nebulae, supernova remnants, or follow-up observations of {\em Swift} or {\em Fermi} sources.

\begin{deluxetable*}{ccccccc}[!htb]
\tablenum{1}
\tablecaption{{\em Chandra} Observations Containing Stellar Bow Shocks\label{table:observation_log}}
\tablehead{
\colhead{}			& \colhead{Obs. Date}	& \colhead{}		& \multicolumn{2}{c}{Nominal Aim Point (J2000)} 		& \colhead{Effective Exp.}	& \colhead{\# Bow}	 \\ \cline{4-5}
\colhead{Obs. ID}	& \colhead{(YYYY-MM-DD)} & \colhead{ACIS-}	& \colhead{R.A.}	& \colhead{Decl.}	& \colhead{Time (ks)}	& \colhead{Shocks}}
\colnumbers
\startdata
748			& 2000-10-15		& S	& 18:46:24.7	& -02:58:34.0	& 34.0		& 1 \\
972			& 2002-03-02		& I	& 18:20:29.9	& -16:10:45.5	& 39.4		& 3	\\
1985		& 2001-07-26		& I	& 14:12:08.0	& -61:45:29.0	& 9.3		& 1	\\
2298		& 2001-05-20		& I	& 18:43:32.1	& -03:54:44.8	& 88.7		& 2	\\
3501		& 2003-08-23		& I	& 10:24:02.5	& -57:45:23.0	& 34.7		& 3	\\
3811		& 2003-10-04		& S	& 12:26:52.8	& -62:49:07.0	& 3.0		& 1	\\
3854		& 2003-07-15		& S	& 19:13:20.3	& +10:11:23.0	& 19.6		& 1	\\
4495		& 2004-09-21		& I	& 10:43:57.5	& -59:32:53.0	& 57.1		& 1	\\
4550		& 2004-07-03		& S	& 17:54:28.3	& -26:20:35.0	& 16.8		& 1	\\
4600		& 2004-07-09		& I	& 18:25:27.0	& -14:48:38.0	& 11.0		& 1	\\
\enddata
\tablecomments{Table~\ref{table:observation_log} is published in its entirety in machine-readable format available from the journal. A portion is shown here for guidance regarding its form and content.}
\end{deluxetable*}

\subsection{Data Reduction and Reprocessing}
\begin{figure}
\includegraphics[width=1\linewidth,clip,trim=2cm 12.5cm 2cm 3cm]{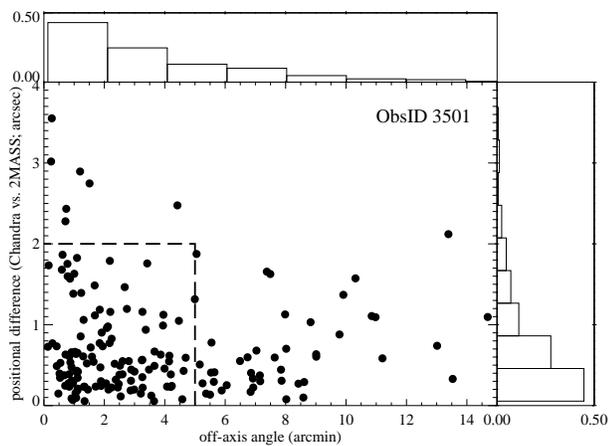}
\caption{The positional offset between the {\em Chandra} X-ray source positions and the matched 2MASS sources, as a function of {\em Chandra} off-axis angle, using ObsID 3501 as an example. The top and right panels show histogram distributions of the off-axis angles (in arcminutes) and positional differences (in arcseconds), respectively. Only sources within the dashed-line box are retained in our analysis. \label{figure:posdiff_offaxis}}
\end{figure}

We used CIAO v4.10 and CALDB 4.7.8 to reprocess all the observations using the \texttt{chandra\_repro} task using standard reduction procedures\footnote{See \url{http://asc.harvard.edu/ciao/threads/index.html}.}. Point source detection was performed using the CIAO task \texttt{wavdetect} \citep{Freeman+02} with scales of 1$^{\prime\prime}$, 2$^{\prime\prime}$, and 4$^{\prime\prime}$ on the 0.5-7 keV image. The absolute astrometric accuracy of {\em Chandra} observations\footnote{See \url{http://cxc.harvard.edu/cal/ASPECT/celmon/}} is typically $\sim$0.8$^{\prime\prime}$. We compare the X-ray point source positions to 2MASS positions and use CIAO task \texttt{reproject\_aspect} to refine the positional accuracy of our X-ray images. Figure~\ref{figure:posdiff_offaxis} shows the positional difference between {\em Chandra} and 2MASS sources as a function of off-axis angle for Obs ID 3501, a particularly crowded field; only sources within the dashed-line box (defined as off-axis angles $<5^{\prime}$ and positional difference $<2^{\prime\prime}$) are retained in our analysis.

All X-ray sources in our preliminary list were then masked out of the X-ray images. We inspected each observation for background flares using the \texttt{lc\_clean} script; background light curves were clipped at 5$\sigma$ to create good-time intervals. Each observation was then filtered using the new good time intervals (e.g., the ``effective'' exposure time listed in Table~\ref{table:observation_log}), and we restricted the energy range for each observation to 0.5-7 keV. We then use the CIAO task \texttt{fluximage} to create exposure maps and exposure-corrected images of each IR bow shock region.  The cleaned, energy-restricted event files were used for the remainder of our analysis.

\begin{figure}
\includegraphics[width=1\linewidth,clip,trim=4cm 12.9cm 3cm 4cm]{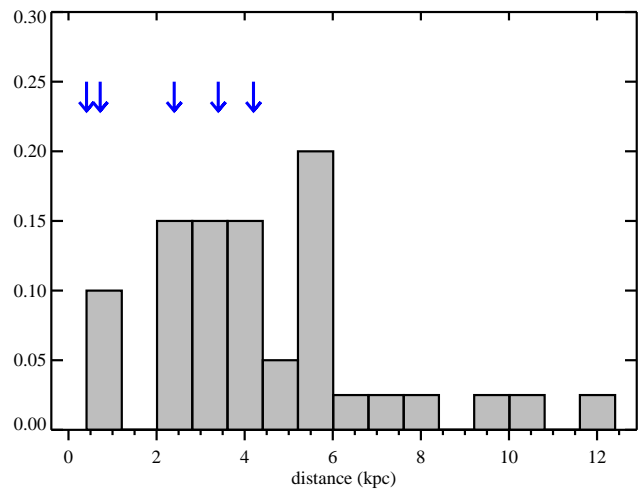}
\caption{Histogram of distances of X-ray detected FOV massive stars (black). The distances of the 5 bow shock driving stars with reliable parallaxes are indicated with downward blue arrows. FOV stars span a much larger range in distances (4.5$\pm$1.0 kpc) than driving stars (2.3$\pm$0.5 kpc).  \label{figure:distance_histogram}}
\end{figure}

To create our final X-ray source lists, only sources with $>$5 counts within 5$^{\prime}$ of the nominal aim point were preserved. The sources were then visually examined for possible false detections. The positions of the detected X-ray sources were checked against the position of the bow shock driving stars and other known OB stars using SIMBAD. We derived distances to the bow shock driving stars and other known OB stars in the field of view (hereafter referred to as ``FOV stars'') using {\em Gaia} DR2 parallaxes \citep{GaiaDR2}. Figure~\ref{figure:distance_histogram} shows the distribution of distances for driving stars and FOV stars with reasonably small distance uncertainties (i.e., $d/\sigma>3$). We find that bow shock driving stars are systematically closer (2.3$\pm$0.5 kpc) than FOV stars (4.5$\pm$1.0 kpc). Table~\ref{table:X-ray_star_properties} provides a summary of all the X-ray detected OB stars in our study, in order of decreasing 0.5-7 keV counts; our procedure for estimating the star's X-ray luminosity is described in Section~\ref{section:stars}. Only 5 out of the 65 unique bow shocks ($\sim$8\%) contained within a {\em Chandra} observation had their driving star detected with $>$5 counts.

\begin{table*}
\tablenum{2}
\caption{Bow Shocks in Sample}\label{table:X-ray_bowshock_properties} 
\begin{center}
\scriptsize
\begin{tabular}{ccccccc}
\hline \hline
			& RA		& Decl.		& $R_0$					& P.A.          & X-ray Exp.	& Off-Axis					        	 \\ \cline{2-3}
Name		& (J2000)	& (J2000)	& ($^{\prime\prime}$)	& ($^{\circ}$)  & Time (ks)		& Angle\tablenotemark{a} ($^{\prime}$)	\\
(1)			& (2)		& (3)		& (4)					& (5)			& (6)           & (7)	\\
\hline
G015.0749-00.6461	& 18:20:22.8	& -16:08:34	& 2.5	& 140   & 325.9	& 2.2		\\
G015.0812-00.6570	& 18:20:25.9	& -16:08:32	& 14.5	& 135   & 325.9	& 2.2		\\
G015.1032-00.6489	& 18:20:26.7	& -16:07:09	& 6.1	& 150   & 325.9	& 3.6		\\
G338.2791-00.01416	& 16:40:33.3	& -46:33:20	& 18.4	& 110   & 205.6	& 1.8		\\
G029.7810-00.2176	& 18:46:27.1	& -02:54:12	& 6.7	& 145   & 183.7	& 4.4		\\
G172.0813-02.2592	& 05:16:18.2	& 34:18:44	& 9.9	& 350   & 140.5	& 0.1		\\
G284.2999-00.3359	& 10:24:11.5	& -57:46:47	& 3.9	& 295   & 135.0	& 1.4		\\
G284.3011-00.3712	& 10:24:03.1	& -57:48:36	& 2.9	& 135   & 135.0 & 3.2		\\
G327.2741-00.14811	& 15:51:10.3	& -54:17:46	& 15.8	& 100   & 106.1	& 0.7		\\
G028.4787-00.0056	& 18:43:19.0	& -03:57:54	& 8.2	& 305   & 98.7	& 3.2		\\
G028.5028-00.0343	& 18:43:28.0	& -03:57:24	& 3.1	& 25    & 98.7	& 2.7		\\
G359.0835-00.4367	& 17:45:08.2	& -29:56:45	& 26.9	& 160   & 89.4	& 0.5   	\\
G041.5496+00.0975	& 19:06:54.0	& 07:42:30	& 6.5	& 115   & 87.9	& 4.1		\\
G348.2233+00.46284	& 17:12:20.8	& -38:29:31	& 40.0	& 160   & 76.8	& 0.1		\\
G006.2812+23.5877	& 16:37:09.5	& -10:34:02	& 29.0	& 30    & 72.1	& 2.8 	    \\
G014.4703-00.6427	& 18:19:10.6	& -16:40:27	& 14.7	& 210   & 66.0	& 2.5		\\
G287.1148-01.0236	& 10:40:12.4	& -59:48:10	& 15.8	& 285   & 59.6	& 1.4		\\
G288.1505-00.5059	& 10:49:25.0	& -59:49:44	& 4.2	& 220   & 59.5	& 4.3		\\
G287.4389-00.6132	& 10:44:00.9	& -59:35:46	& 4.0	& 325   & 57.3	& 2.9		\\
G287.6736-01.0093	& 10:44:11.1	& -60:03:21	& 3.7	& 5     & 56.5	& 0.9		\\
G080.3667+00.4209	& 20:35:16.1	& 41:12:34	& 17.6	& 125   & 56.3	& 4.9		\\
G287.4071-00.3593	& 10:44:43.9	& -59:21:25	& 7.8	& 110   & 55.3	& 0.1		\\
G333.3639-00.3327	& 16:21:37.7	& -50:21:17	& 5.3	& 30    & 54.3	& 3.0		\\
G333.3665-00.4112	& 16:21:59.3	& -50:24:31	& 40.6	& 160   & 54.3	& 0.5		\\
G353.0649+01.2985	& 17:22:50.0	& -34:03:22	& 10.2	& 25    & 39.5	& 1.3		\\
G012.8263-00.1278	& 18:14:0.3		& -17:52:32	& 7.5	& 355   & 38.2	& 3.2		\\
G012.8724-00.23427	& 18:14:29.4	& -17:53:10	& 31.0	& 130   & 38.2	& 2.6		\\
G026.1437-00.0420	& 18:39:09.1	& -06:03:28	& 22.1	& 50    & 37.5	& 1.7		\\
G284.3400-00.2827	& 10:24:39.2	& -57:45:21	& 4.1	& 285   & 36.2	& 0.6		\\
G045.2527-00.4751	& 19:15:54.0	& 10:43:36	& 5.4	& 14    & 29.6	& 2.4		\\
G016.8993-01.1152	& 18:25:38.9	& -14:45:06	& 24.0	& 120   & 29.0	& 3.5		\\
G026.6435-00.0209	& 18:40:06.1	& -05:37:03	& 6.8	& 305   & 20.1	& 3.4		\\
G044.5388-00.2303	& 19:13:40.0	& 10:12:30	& 4.5	& 225   & 19.6	& 1.2		\\
G337.1246-00.1209	& 16:36:29.9	& -47:29:11	& 6.6	& 225   & 19.5	& 4.3		\\
G003.2532-00.2836	& 17:54:15.6	& -26:17:24	& 25.3	& 290   & 19.1	& 3.2		\\
G337.2776-00.2660	& 16:37:44.5	& -47:28:13	& 10.9	& 30    & 19.0	& 3.7		\\
G336.9225+00.05767	& 16:34:54.6	& -47:30:54	& 11.0	& 270   & 19.0	& 3.3		\\
G022.4119-00.2010	& 18:32:48.0	& -9:26:40	& 10.1	& 55    & 18.2  & 4.8		\\
G021.6949+00.24498	& 18:29:51.0	& -09:52:26	& 26.5	& 265   & 15.2	& 1.0		\\
G000.1169-00.5703	& 17:48:07.0	& -29:07:56	& 26.4	& 27    & 14.1	& 2.4		\\
G033.8783+00.2033	& 18:52:26.0	& 00:56:06	& 14.0	& 155   & 11.9	& 4.3		\\
G028.8142+00.1922	& 18:43:13.1	& -03:34:34	& 3.9	& 230   & 10.9	& 2.8		\\
G347.8625-00.1941	& 17:13:59.4	& -39:10:11	& 4.4	& 240   & 10.7	& 2.1		\\
G046.7943-00.2716	& 19:18:06.0	& 12:11:06	& 19.6	& 130   & 9.9	& 1.6		\\
G319.9407+00.2825	& 15:05:28.1	& -58:04:54	& 2.5	& 240   & 9.9	& 0.6		\\
G312.2963-00.42721	& 14:12:03.9	& -61:49:27	& 17.5	& 90    & 9.3	& 4.0		\\
G003.5118-00.0470	& 17:53:56.0	& -25:56:50	& 8.6	& 135   & 6.8	& 3.6		\\
G332.9033+00.11195	& 16:17:36.9	& -50:21:41	& 16.4	& 50	& 5.0   & 1.5		\\
G031.9308+00.2676	& 18:48:39.1	& 00:46:08	& 15.5	& 10    & 4.7	& 0.6		\\
G031.9747+00.3348	& 18:48:30.0	& 00:41:57	& 6.9	& 230   & 4.7	& 3.6		\\
G356.6602+00.9209	& 17:33:47.9	& -31:16:27	& 15.8	& 105   & 4.6	& 4.0		\\
G028.6268-00.16592	& 18:44:09.9	& -03:54:24	& 13.3	& 335   & 3.1	& 1.2		\\
G300.1020-00.1371	& 12:26:34.4	& -62:52:20	& 1.5	& 100   & 3.0	& 3.2		\\
G021.6408-00.18419	& 18:31:17.6	& -10:07:18	& 27.7	& 105   & 2.8	& 1.3		\\
G345.3932-00.0552	& 17:05:45.7	& -41:04:17	& 31.0	& 210   & 2.4	& 0.7		\\
G352.4288-00.0241	& 17:26:24.9	& -35:19:34	& 11.9	& 220   & 2.3	& 2.8		\\
G352.4634+00.0358	& 17:26:16.0	& -35:15:50	& 7.4	& 130   & 2.3	& 1.0		\\
G003.7391+00.1425	& 17:53:43.0	& -25:39:19	& 3.9	& 20    & 2.0	& 0.7    	\\
G341.3576+00.1490	& 16:51:14.4	& -44:06:37	& 5.0	& 225   & 1.8	& 3.1		\\
G343.9419-00.1440	& 17:01:22.1	& -42:16:40	& 4.2	& 210   & 1.3	& 4.5		\\
G060.9223-00.1176	& 19:46:22.7	& 24:37:48	& 4.9	& 195   & 1.3	& 1.0		\\
\hline \hline
\end{tabular}
\end{center}
\tablecomments{$^a$For sources with multiple ObsID, only the largest off-axis angle is reported. } 
\end{table*}

\begin{table*}
\tablenum{3}
\caption{X-ray Detected OB Stars}\label{table:X-ray_star_properties} 
\begin{center}
\tiny
\setlength\tabcolsep{2.5pt}
\begin{tabular}{cccccccccc}
\hline \hline
R.A.    & Decl.							& Stellar		& Spectral	& Distance\tablenotemark{b}	& Net Cts/Exp. Time	& Off-Axis			& log$L_X$ & Driving	& 		\\ \cline{1-2}
\multicolumn{2}{c}{(X-ray source; J2000)}	& Counterpart	& Type\tablenotemark{a}		& (kpc)		& (0.5-7 keV)/(ks)			& Angle ($^\prime$)\tablenotemark{c}	& (erg s$^{-1}$) & Star?		& Notes\tablenotemark{d}	\\
\hline
05:16:18.1 & +34:18:44.3	& HD 34078						& O9.5V							& 0.41$\pm$0.01	& 3004/140.5 & 0.1  & 31.1  & Y	& M	\\
17:12:20.9 & -38:29:30.4	& CD-38 11636					& O8							& 0.72$\pm$0.25	& 974/78.8	& 0.1   & 31.3  & Y	& M	\\
10:27:58.0 & -57:45:49.0	& V* V712 Car  					& O3If*/WN6+O3If*/WN6			& 5.3$\pm$1.4	& 660/34.7	& 1.4   & 33.2  & N	& 	\\
10:24:01.2 & -57:45:31.0	& Wd~2 MSP 188					& O5.5, O3V+O5.5				& 5.3$\pm$1.4	& 527/34.7	& 0.6   & 33.1  & N	& C	\\
10:24:18.4 & -57:48:29.5	& WR~20b 						& WN6ha							& 3.6$\pm$0.8	& 408/34.7	& 4.9	& 32.7  & N	& 	\\
10:24:02.4 & -57:44:35.9	& Wd~2 5						& O5/5.5V/III(f) 				& 5.3$\pm$1.4	& 337/34.7	& 0.8   & 32.9  & N	&	\\
10:24:01.9 & -57:45:27.7	& Wd~2 MSP 167 					& O8V, O8V						& 4.2$\pm$0.5	& 264/34.7	& 0.4   & 32.6  & N	& C	\\
10:44:43.9 & -59:21:25.1	& HD 93249   					& O9III							& 3.4$\pm$0.4	& 209/55.3	& 0.1   & 32.1  & Y	& PM	\\
10:40:31.7 & -59:46:43.9	& HD 92644 						& B0V							& 2.6$\pm$0.4	& 139/59.1	& 2.8   & 31.7  & N	& M	\\
10:44:11.1 & -60:03:21.5	& HD 305536						& O9.5V							& 2.4$\pm$0.2	& 126/56.5	& 4.9   & 31.6  & Y	&	\\
10:24:02.2 & -57:45:31.3	& [RSN2011] C, D, E				& O7V, O9.5, O6-7				& 3.1$\pm$0.8	& 112/34.7	& 0.3   & 32.0  & N	& C	\\
18:31:16.5 & -10:09:24.8	& UCAC 160-180469				& WR8							& $<$9.9		& 82/2.8	& 0.8   & \nodata & N	&	\\	
10:24:00.2 & -57:45:32.4	& 2MASS J10240020-5745327		& O8.5V							& 5.7$\pm$2.4	& 68/34.7	& 0.8   & 32.3  & N	&	\\	
10:24:00.7 & -57:49:25.2	& 2MASS J10240073-5745253 		& O8V, O6.5V					& 6.7$\pm$1.3	& 68/34.7	& 0.7   & 32.4  & N	& C	\\	
10:23:55.1 & -57:49:26.6	& Wd~2 MSP 165					& O4V							& 6.0$\pm$1.7	& 63/34.7	& 2.1   & 32.3  & N	&	\\	
10:24:16.2 & -57:43:43.7	& Wd~2 NRM 2					& O8.5III						& 5.9$\pm$1.5	& 57/34.7	& 3.6   & 32.3  & N	&	\\	
10:4405.9 & -59.9948782	    & HD 305520						& B1Ib							& 2.3$\pm$0.2	& 45/56.5	& 1.7   & 31.1  & N	&	\\	
10:44:42.1 & -59:22:30.6	& CPD-58 2655 					& B1V							& 2.6$\pm$0.2	& 42/55.3	& 1.2   & 31.2  & N	&	\\	
18:13:58.2 & -17:56:25.4	& [MCF2015] 6, 7				& B0-5, O4-6					& 4.0$\pm$1.8	& 39/38.2	& 3.9   & 31.7  & N	& C	\\	
10:44:35.9 & -59:23:35.6	& Tr~15 C28						& O9V							& 2.6$\pm$0.2	& 28/55.3	& 3.0   & 31.0  & N	&	\\	
10:23:56.1 & -57:45:29.9	& Wd~2 MSP 182					& O4III							& 5.7$\pm$1.2	& 27/34.7	& 1.8   & 31.9  & N	& 	\\	
10:24:01.5 & -57:45:56.9	& Wd~2 MSP 263					& O6V							& $<$10.2		& 25/34.7	& 0.7   & \nodata & N &	\\	
10:24:04.8 & -57:45:28.1	& Wd~2 MSP 171					& O5V							& 4.0$\pm$0.6	& 22/34.7	& 0.4   & 31.5  & N	&	\\	
10:44:43.7 & -59:21:17.3	& Tr~15 2  						& O9/9.5III						& 3.0$\pm$0.3	& 19/55.3	& 0.1   & 31.0  & N	&	\\	
10:48:58.9 & -59:41:09.2	& HD 303413						& B1Ib							& 2.8$\pm$0.2	& 15/50.0	& 5.0   & 30.9  & N	&	\\	
10:44:42.3 & -59:23:03.8	& CPD-58 2656					& B0.5IV/V						& 3.1$\pm$0.3	& 15/55.3	& 1.7   & 30.9  & N	&	\\	
10:24:02.4 & -57:45:46.8	& Wd~2 MSP 235					& O9.5V							& 3.8$\pm$0.8	& 15/34.7	& 0.5   & 31.3  & N	&	\\	
10:24:00.9 & -57:45:05.0	& Wd~2 MSP 96					& B1+B1							& 4.1$\pm$0.7	& 14/34.7	& 0.7   & 31.3  & N	&	\\	
10:24:01.3 & -57:45:29.5	& Wd~2 MSP 175 					& O4V							& 4.2$\pm$0.5	& 13/34.7	& 0.5   & 31.3  & N	&	\\	
10:24:00.4 & -57:44:44.2	& Wd~2 MSP 44					& B1V+PMS						& 7.0$\pm$2.0	& 13/34.7	& 1.0   & 31.8  & N	&	\\	
17:05:45.6 & -41:04:17.3	& TYC 7873-1475-1				& OB+							& 4.2$\pm$2.2	& 12/2.4	& 2.6   & 32.4  & Y	&	\\	
18:13:59.7 & -17:57:41.2	& [MCF2015] 8					& O4-6							& \nodata		& 11.38.2	& 4.0   & \nodata & N	&	\\	
12:26:36.7 & -62:47:52.0	& CD-62 653						& B7e							& 1.2$\pm$0.1	& 10/3.0	& 4.0   & 31.2  & N	&	\\	
10:24:00.3 & -57:45:42.5	& 2MASS J10240034-5745426		& O8V							& 0.69$\pm$0.22	& 9/34.7	& 0.8   & 29.6  & N	& C	\\	
18:14:20.5 & -17:56:11.2	& [MCF2015] 23 					& O6-7							& \nodata		& 9/38.2	& 1.8   & \nodata & N	&	\\	
10:49:04.4 & -59:48:00.2	& LS 1940						& OB							& 2.3$\pm$0.2	& 7/50.0	& 2.9   & 30.4  & N	&	\\	
10:44:46.5 & -59:21:53.9	& CPD-58 2662 					& B2V							& 3.2$\pm$0.3	& 7/55.3	& 0.8   & 30.6  & N	&	\\	
10:40:23.4 & -59:50:39.0	& [AHP2016] OBc 3          		& O7V							& 11.7$\pm$4.9	& 7/26.8	& 2.9   & 32.1  & N	&	\\	
10:24:06.6 & -57:47:15.7	& Wd~2 NRM 3					& O9.5V							& 5.1$\pm$1.4	& 6/34.7	& 2.0	& 31.2  & N	&	\\	
10:23:57.7 & -57:45:34.2	& Wd~2 MSP 201 					& OB							& 8.0$\pm$3.2	& 5/34.7	& 1.4   & 31.5  & N	&	\\	
\hline \hline
\end{tabular}
\end{center}
\tablecomments{Known X-ray binaries were excluded from our sample. 
\tablenotetext{a}{Spectral types were taken from SIMBAD.}
\tablenotetext{b}{Distances were derived using the {\em Gaia} second data release \citep{GaiaDR2}.} 
\tablenotetext{c}{For sources detected in multiple ObsIDs, we report the largest off-axis angle.}
\tablenotetext{d}{Explanation of flags: M - multiple obs IDs; C - counterpart confusion; PM - high proper motion}
}
\end{table*}

	\subsection{Bow Shock ``Postage Stamps''}
We created ``postage stamp'' {\em Chandra} images for each of the 60 unique bow shocks from the exposure-corrected ``fluxed'' images. We masked the location of the driving star for each bow shock with a circular region that was approximately twice the radius of the PSF full width at half maximum at the detector location, so that our resulting stacked image did not contain low-level X-ray emission from the driving stars. Generally, the size of the {\em Chandra} PSF is less than the stand-off radius $R_0$ between the driving star and the bow shock \citep[see, e.g.,][their Appendix A for details on modeling the {\em Chandra} PSF]{Primini+11}, so inadvertent masking of X-ray counts that are genuinely associated with the IR bow shock location is unlikely.  The size of the box was set to three times the stand-off radius ($R_0$) of the bow shock. Bow shocks in the \citet{Kobulnicky+16} catalog have measured position angles giving the orientation of the IR bow shock relative to Galactic north; we use the same methodology to measure the position angles of the bow shocks in the Milky Way Project using the 24 \micron\ images.

\begin{figure*}
\begin{center}
\includegraphics[width=0.85\linewidth,clip,trim=1cm 10cm 1cm 9.4cm]{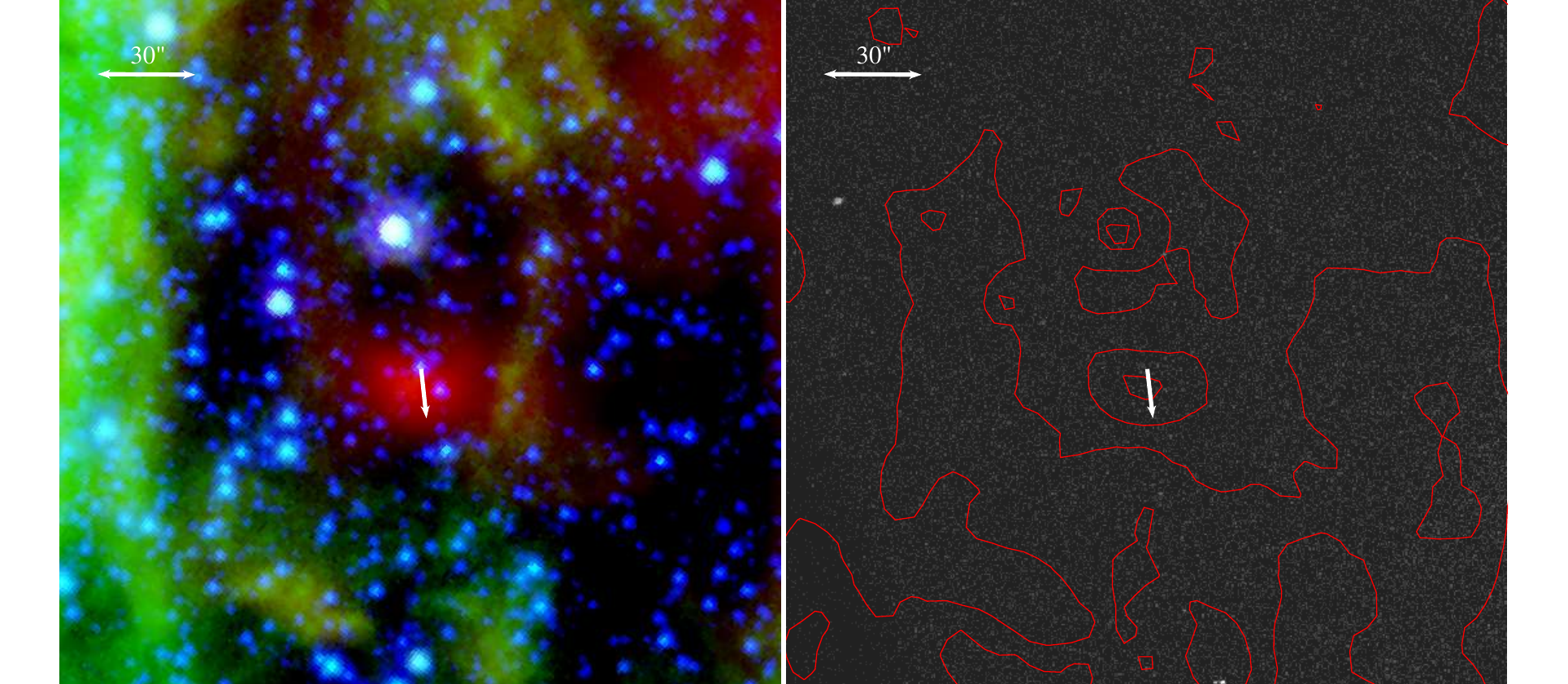} \\
\includegraphics[width=0.85\linewidth,clip,trim=1cm 10cm 1cm 9.4cm]{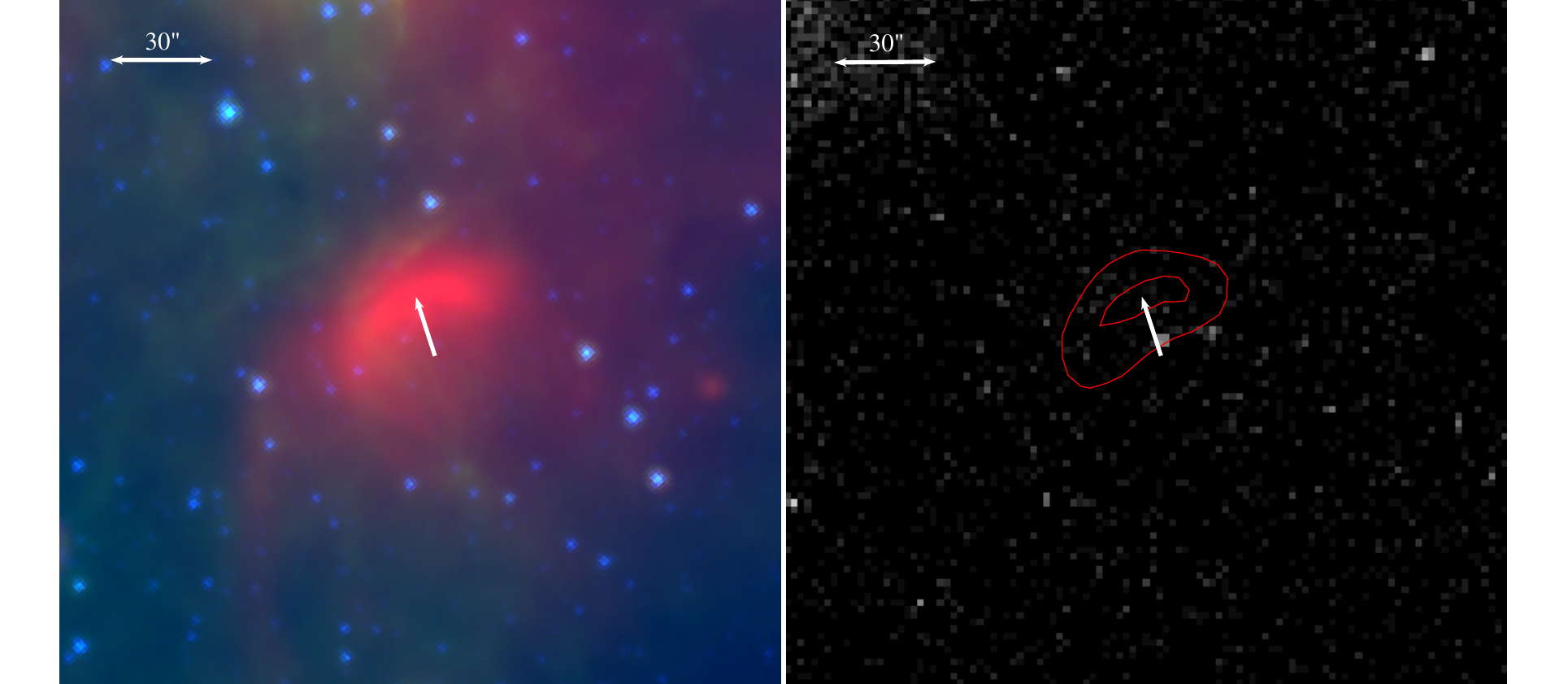}	
\end{center}
\caption{{\it Left}: Infrared RGB images of the MWP bow shocks G327.2741-00.14811 (top) and G338.2791-00.01416 (bottom; Jayasinghe et al. \textit{in prep}), with the standoff radii $R_0$ and position angles indicated by the white vector. Red is MIPS 24 $\mu$m, green is IRAC4 (8 $\mu$m), and blue is IRAC1 (3.6 $\mu$m). {\it Right}: The merged 0.5-7 keV images of the same region; red lines show the 24 $\mu$m contours. All images are displaying in logarithmic scaling; the IR images are in units of MJy/sr and the X-ray images are in counts. \label{figure:XrayIR_bowshock_examples}}
\end{figure*}

For each image, we define a vector located at the coordinates of the driving star, with a magnitude (length) $R_0$ (the standoff radius) that is pointed at the position angle of the IR bow shock. Figure~\ref{figure:XrayIR_bowshock_examples} show two example IR bow shocks from the Milky Way Project with deep ($>$100 ks) corresponding X-ray image and bow shock vectors.

\section{Bow Shock Stacking Analysis}\label{section:bow_shocks_stacking}
To create the final stacked X-ray image of the bow shocks, we used the CIAO task \texttt{dmregrid} to rotate and resize each image vector so that the IR bow shock is located at an ($x,y$) position (0.5$n$,0.67$n$), where $n$ is the size of the rescaled image. We tested several final image sizes (50, 70, and 100 pixels on a side) but our choice of final image size did not affect our results. 

Our stacked X-ray image\deleted{s} reveals an average photon flux at the location of the IR bow shock of $\sim$1.1$\times10^{-7}$ ph s$^{-1}$. We use the CIAO task \texttt{modelflux} to convert this photon flux to an energy flux, assuming an absorbed power law spectral model ($\Gamma=2$) and a typical Milky Way absorbing column of $\sim6\times10^{21}$ cm$^{-1}$. The predicted 0.5-7 keV energy flux is 4.4$\times10^{-16}$ \flux\ (unabsorbed). Assuming an average distance to a Galactic bow shock of $\sim$2 kpc, this flux corresponds to a total unabsorbed 0.5-7 keV luminosity of $\sim$2.2$\times10^{29}$ \lum. For comparison, the predicted ``background'' flux for this image (selected from a region in the stacked image that is not expected to contain any X-ray sources) is 3.9$\times10^{-16}$ \flux\ (corresponding to a flux of 1.9$\times10^{29}$ \lum\ at a distance of 2 kpc).

Although $\sim$70\% of the IR bow shocks in the \citet{Kobulnicky+16} catalog are in ``isolated'' environments, $\sim$34\% of the total X-ray exposure time in our sample was targeted at star-forming regions that likely contain significant diffuse X-ray emission. Spectral fitting of {\em Chandra} observations of five massive star-forming regions \citep{Townsley+11} found the diffuse emission to be dominated by at least two thermal components, one at $kT\sim$0.2-0.6 keV and one with $kT\sim$0.5-0.9 keV, with an integrated surface brightness of $\sim(3-500)\times10^{30}$ erg s$^{-1}$ pc$^{-2}$. The physical size of a ``postage stamp'' that is 70 pixels on a side (with a plate scale of 0.492$^{\prime\prime}$ per pixel) at an average distance of 2 kpc is $\sim$0.32 pc on a side ($\sim$0.10 pc$^{2}$). The total X-ray luminosity in each ``postage stamp'' is therefore expected to be (0.3--50)$\times10^{29}$ \lum, in agreement with our background estimate above. We therefore do not find any evidence of excess X-ray emission in our stacked image that can be attributed to stellar wind bow shocks.

\begin{figure*}
\begin{center}
\includegraphics[width=0.8\linewidth,clip,trim=2cm 9.7cm 2cm 9.5cm]{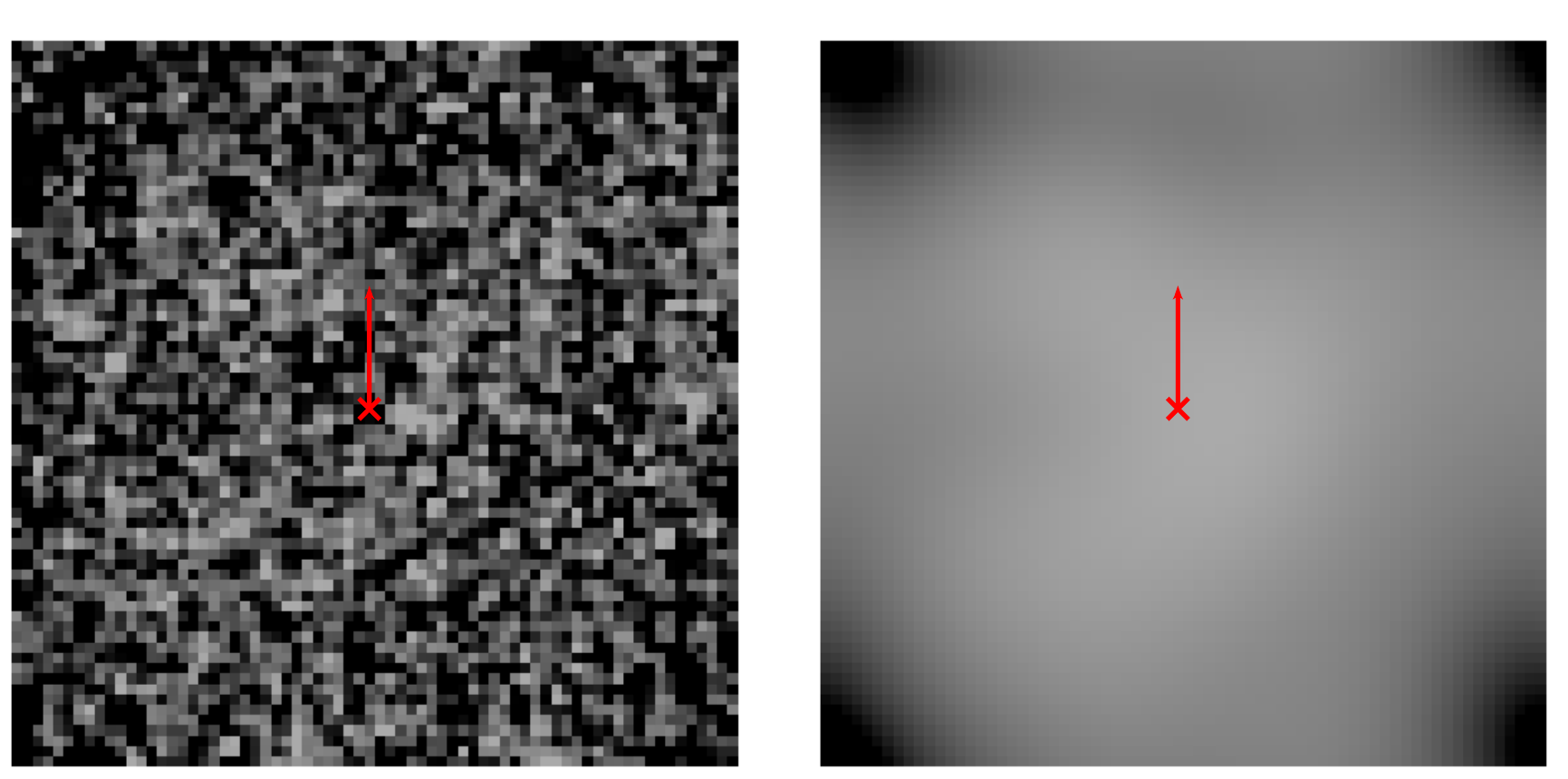}
\end{center}
\caption{The stacked 70 pixel $\times$ 70 pixel 0.5-7 keV bow shock image in photon flux units (ph s$^{-1}$ cm$^{-2}$)}, constructed from 61 individual {\em Chandra} exposures totaling 1.73 Msec for bow shocks in isolated environments. The raw figure is shown on the left, while the right panel has been smoothed for display purposes only. The red ``X'' shows the location of the central driving stars and the red arrow indicates the location of the IR bow shocks. No excess X-ray emission is detected at the location of the IR bow shocks.\label{figure:stacked_bowshock}
\end{figure*}

We perform an additional stacking analysis using only the 42 bow shocks from the \citet{Kobulnicky+16} in ``isolated'' environments and the 19 bow shocks from the Milky Way Project that were not associated with a Galactic bubble. Using only the ``isolated'' bow shocks ($\sim$67\% of our sample by number) yields a total exposure time of 1.73 Msec. We follow the same stacking analysis as for our full sample; at the location of the IR bow shocks the average photon flux is 1.2$\times10^{-7}$, yielding an energy flux of 4.9$\times10^{-16}$ \flux\ ($\sim$2.4$\times10^{29}$ \lum\ at 2 kpc). The predicted background flux is consistent with the full bow shock sample; we again find no evidence for excess X-ray emission at the location of the IR bow shock. Figure~\ref{figure:stacked_bowshock} show the stacked X-ray image of the isolated bow shocks (both in ``raw'' photon counts, and an image that has been smoothed for display purposes); the full sample images look similar.

\section{X-ray Characteristics of Bow Shock Driving Stars}\label{section:stars}
We additionally investigated whether the characteristics of X-ray detected bow shock driving stars differed significantly from X-ray detected FOV stars. The X-ray spectra of many of the massive stars listed in Table~\ref{table:X-ray_star_properties} have been previously analyzed in the literature; here, we only aim to approximate their X-ray luminosities assuming a simplified X-ray spectral model. 

We assume a typical X-ray spectrum composed of an absorbed thermal plasma (\texttt{tbabs*apec} in \texttt{XSPEC}), with $N_{\rm H}\sim6\times10^{21}$ cm$^{-2}$ and a plasma temperature of $\sim$0.9 keV \citep[see, e.g.][for a detailed study of X-ray spectra of massive stars with {\em XMM-Newton}]{Naze09}, and use \texttt{PIMMS}\footnote{See \url{https://heasarc.gsfc.nasa.gov/docs/software/tools/pimms.html}} to convert the observed count rate for each source into a 0.5-7 keV flux. This flux is then converted into a luminosity for stars with {\em Gaia} distance estimates. In Table~\ref{table:LX_compared_to_literature}, we compare our estimated luminosities for four well-studied massive stars to the values available in the literature. The most significant deviations occur where the {\em Gaia} distances disagree with those assumed in earlier studies; we therefore scale the literature values to the {\em Gaia} distances for a more direct comparison. In general, our approximate luminosities agree with the literature values within a factor of a few. We therefore do not anticipate systematic differences in luminosity across our full massive star sample.

\begin{table}
\tablenum{4}
\caption{X-ray Luminosities of Selected Massive Stars Compared to the Literature}\label{table:LX_compared_to_literature}
\begin{center}
\scriptsize
\begin{tabular}{cccc}
\hline \hline
        & \multicolumn{2}{c}{$L_X$ (erg s$^{-1}$)}   &  \\ \cline{2-3}
Star    & This Work     & Literature\tablenotemark{a}    & Reference    \\
(1)    & (2)           & (3)             & (4)          \\
\hline
V* V712 Car     & 1.6$\times10^{33}$    & 4.0$\times10^{33}$    & \citet{Tsujimoto+07}  \\
WR 20b          & 5.0$\times10^{32}$    & 7.3$\times10^{32}$    & \citet{Tsujimoto+07}  \\
Wd~2 5          & 7.9$\times10^{32}$    & 1.3$\times10^{33}$    & \citet{Tsujimoto+07}  \\ 
HD 305536       & 4.0$\times10^{31}$    & 1.4$\times10^{31}$    & \citet{Broos+11}      \\
\hline \hline
\end{tabular}
\end{center}
\tablecomments{
\tablenotetext{a}{Literature luminosity has been scaled to the distance reported in Table~\ref{table:X-ray_star_properties}. }
}
\end{table}

\begin{figure*}
\begin{tabular}{cc}
\includegraphics[width=0.49\linewidth,clip,trim=4cm 12.7cm 3cm 2cm]{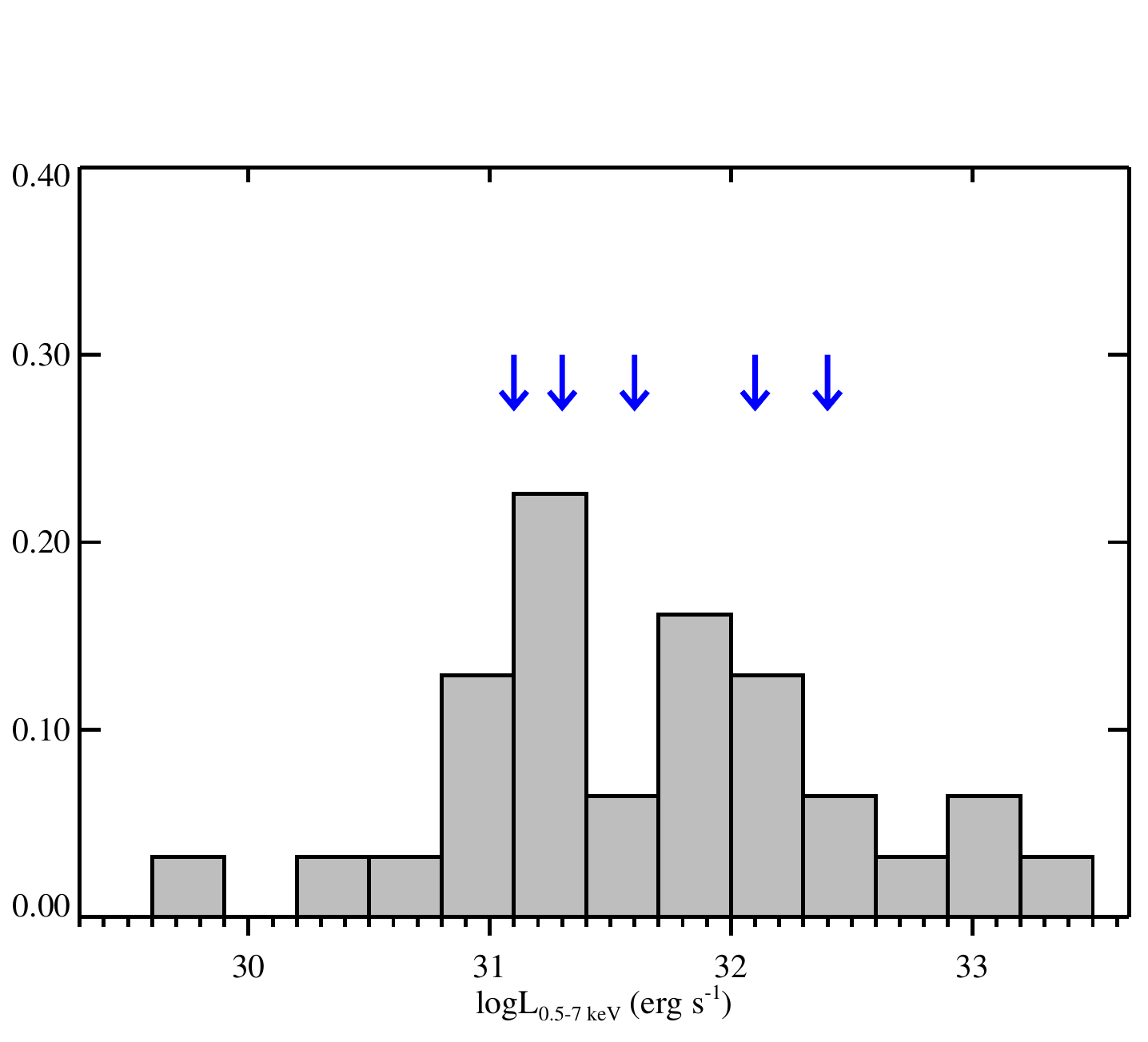} &
\includegraphics[width=0.49\linewidth,clip,trim=4cm 12.7cm 3cm 2cm]{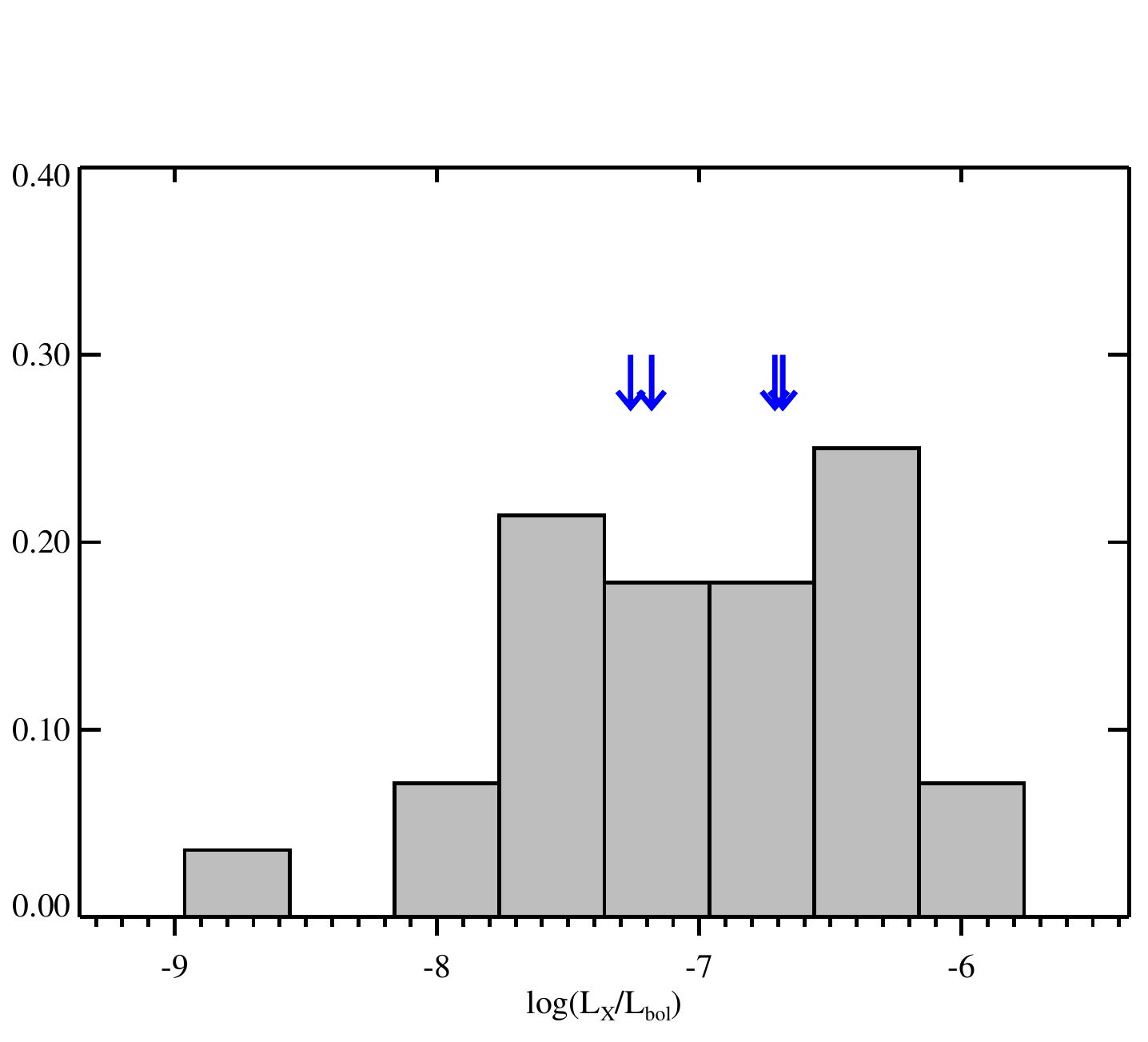}
\end{tabular}
\caption{\textit{Left}: Histogram of X-ray luminosities of FOV stars (black). The luminosities of the bow shock driving stars are indicated with downward blue arrows. \textit{Right}: Histogram of the (logarithmic) $L_X/L_{\rm bol}$ fraction for FOV stars (black). The $L_X/L_{\rm bol}$ ratios of bow shock driving stars with known spectral types are indicated with downward blue arrows. The overall sample average $L_X/L_{\rm bol}\sim1.1\times10^{-7}$. \label{figure:lx_lbol}}
\end{figure*}

We find no evidence of systematic differences in X-ray luminosity between the FOV stars and the five bow shock driving stars. The average 0.5-7 keV luminosity of stars in our sample is $\sim5\times10^{31}$ \lum, while the average bow shock driving star luminosity is $\sim3\times10^{31}$ \lum. We also investigated whether the stars in our sample follow the well-known $L_X \sim10^{-7} L_{\rm bol}$ relationship \citep{Chlebowski89,Berghoefer+97,Naze+11}. We obtain the spectral type of each star from SIMBAD. We adopted model luminosities of massive stars \citep[see][their Table~2]{Crowther07} to infer $L_{\rm bol}$ for different spectral types of stars in our sample. In cases where SIMBAD lists multiple spectral type for a star, we assume the earliest type. Figure~\ref{figure:lx_lbol} shows the distributions of $L_X$ and $L_X/L_{\rm bol}$ ratios for our sample. We find an average $L_X/L_{\rm bol}\sim1.1\times10^{-7}$, in excellent agreement with previous studies. None of the stars in our sample exhibit unusually high $L_X/L_{\rm bol}$ ratios expected from magnetic OB stars \citep{Petit+13,udDoula+16,Gagne+05,Naze+12}. We note that {\em all} of the X-ray detected bow shocks driving stars are of ``late-types'' (e.g., later than O6), while the FOV star sample contains both early- (e.g., earlier than O5.5) and evolved Wolf-Rayet stars. The number of X-ray detected Wolf-Rayet and early-O type stars is small, and exhibit $L_X/L_{\rm bol}$ ratios that are statistically consistent with our sample average. 

\section{Discussion}\label{section:discussion}
Previous attempts to study the X-ray emission from stellar wind bow shocks have yielded only upper limits on the order of $L_X\lesssim10^{30}$ \lum\ using archival {\em XMM-Newton} observations \citep{Toala+17,Toala+16} and dedicated observations \citep{DeBecker+17} of individual bow shocks. Our stacking analysis adds a new, more stringent upper limit, $L_X\lesssim2\times10^{29}$ \lum. 

The non-thermal radiation model developed by \citet{DeBecker+17} assumes X-rays are generated via inverse Compton scattering at a magnetically-confined shock front. The swept-up dust in the bow shock is heated by photons from the driving star, and emits the reprocessed energy in the IR. These IR photons then interact with the energetic electrons at the shock front and are scattered into the X-ray and $\gamma$-ray regime. The IR luminosity therefore some fraction $\chi_{\rm IR}$ of the bolometric. luminosity of the dust at the shock front \citep[e.g.,][]{vanBuren+88}. The predicted X-ray luminosity additionally depends on the magnetic field strength $B$ at the shock front and the power in relativistic electrons (expressed as a fraction $\chi_{\rm rel}$ of the total power available in the bow shock). When upper limits of $\chi_{\rm IR}=\chi_{\rm rel}=1$ and $B=10^{-4}$ G are assumed \citep[measurements of the ambient ISM magnetic field are consistent with $\sim2-6\mu$G,][]{Meyer+17,HarveySmith+11,Fiedler+93,Troland+86}, their model predicts an integrated X-ray luminosity of $\sim4\times10^{29}$ \lum -- a factor of $\sim$2 above our upper limit. Our non-detection indicates that, if X-ray photons are indeed being produced in IR bow shocks, the production mechanism is much less efficient than the models derived by \citet{DeBecker+17}. 

State-of-the art hydrodynamical codes now include particle acceleration, allowing X-ray production in bow shock nebulae to be self-consistently modeled \citep{vanMarle+18}. For example, the detailed, 2D hydrodynamical treatment by \citet{delValle+18}, which includes diffusion of particles and advection of energy out of the bow shock region, predicts X-ray luminosities on the order of a few $\times10^{28}$ \lum\, consistent with our upper limit, with potentially higher luminosities ($\sim10^{30}$ \lum) produced in the $\gamma$-ray regime. Although these models predict higher luminosities in the $\gamma$-ray than the X-ray, systematic searches for $\gamma$-ray emission from bow shocks have similarly yielded only upper limits \citep{Schulz+14,HESS}.

It is unlikely that X-ray emission from stellar wind bow shocks, if it is indeed produced through the mechanisms previously proposed, is detectable by present X-ray telescopes. Even the future {\em Athena} X-ray Observatory \citep{Athena} will not possess the sensitivity and angular resolution required to detect such faint X-ray emission from IR bow shocks.

The proposed NASA X-ray flagship mission {\em Lynx}, however, will have 50$\times$ greater sensitivity than {\em Chandra} with similar angular resolution over a significantly wider field of view\footnote{See \url{https://www.lynxobservatory.com/}}. We use the Simulated Observations of X-ray Sources\footnote{See \url{http://hea-www.cfa.harvard.edu/~jzuhone/soxs/index.html}} (SOXS) package in Python to determine the feasibility of detecting faint X-ray emission from bow shocks with {\em Lynx}.

For consistency with our assumptions in Section~\ref{section:bow_shocks_stacking}, we assume the non-thermal bow shock spectrum follows a power law with $\Gamma=2$ over an energy range of 0.1-10 keV, subject to absorption due to intervening gas and dust in the Milky Way. We used the nearest bow shock in our sample: driving star HD 34078, at 0.41 kpc. The bow shock nebula is formed at $R_0=9.9^{\prime\prime}$ from its driving star, with a column density along the line of sight $N_{\rm H}=1.2\times10^{21}$ cm$^{-2}$. The spectrum is then renormalized so that the 0.5-7 keV flux is just below our {\em Chandra} detection limit at the distance of HD 34078.

Photons are drawn from the underlying spectrum and spatially distributed according to a $\beta$-model surface brightness profile. The X-ray surface brightness $S$ as a function of radius $r$ is given by

\begin{equation}
    S(r) = S_0 \left[1+\frac{r}{r_{\rm c}} \right]^{-3\beta+1/2},
\end{equation}

\noindent where $S_0$ is the core surface brightness, $r_{\rm C}$ is the core radius, and $\beta$ is the slope parameter. We assume $r_{\rm c}\sim3^{\prime\prime}$ and $\beta=1$ (with an ellipticity of 0.3) to approximately match the typical appearance of the 24 \micron\ emission.

After the photons are assigned a sky location, our simulated bow shock is then ``observed'' in the 0.2-7 keV energy range with the {\em Lynx} high-definition X-ray imager (HDXI) using the SOXS instrument simulator for a given exposure time. The instrument simulator additionally adds a Galactic foreground and instrument noise to our image; we refer the reader to the SOXS User's Guide for further details.

\begin{figure*}[!t]
\begin{tabular}{cc}
\includegraphics[width=0.45\linewidth,clip,trim=1cm 5cm 1cm 2cm]{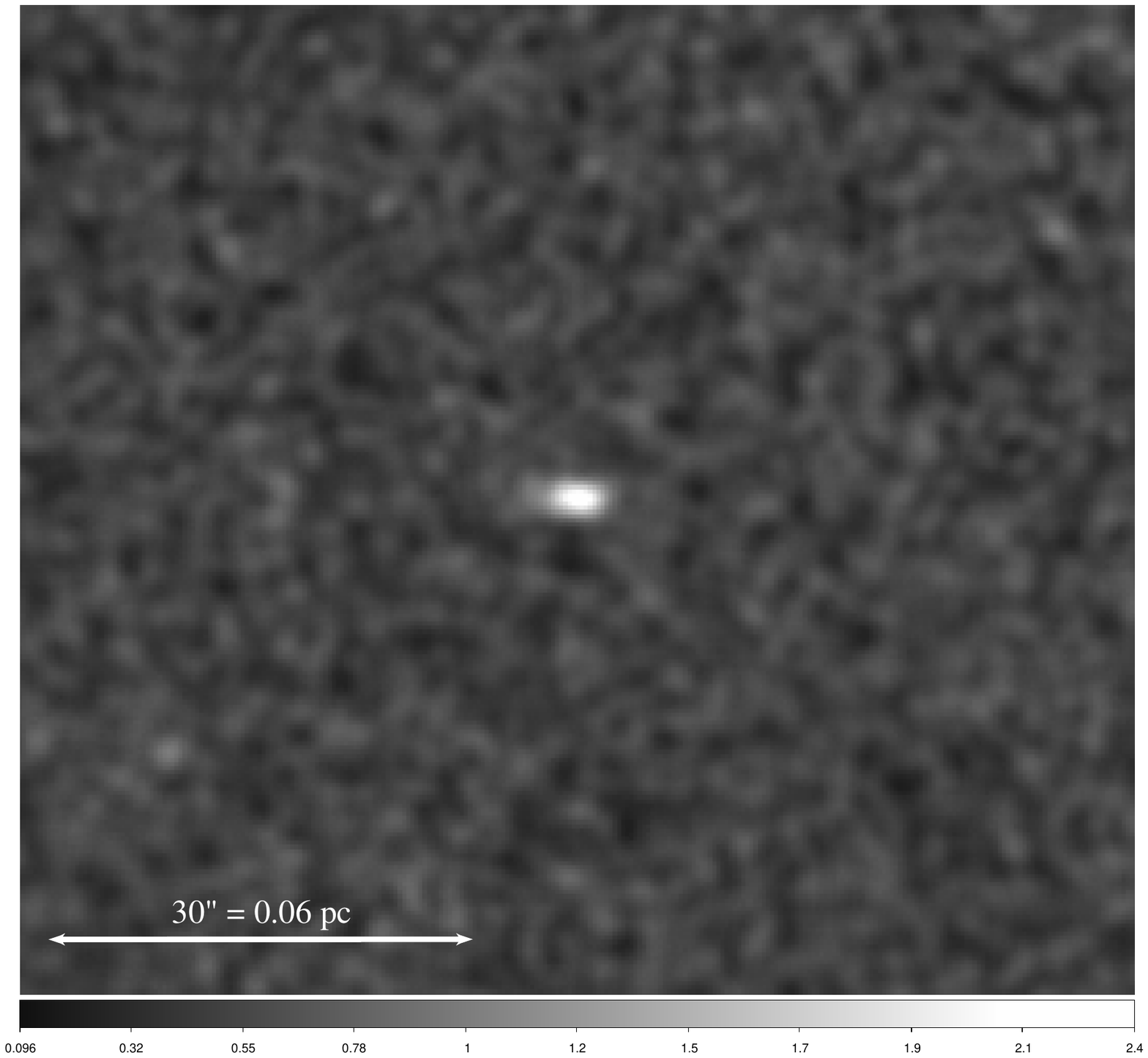} &
\includegraphics[width=0.5\linewidth,clip,trim=2.5cm 12.9cm 2.5cm 3cm]{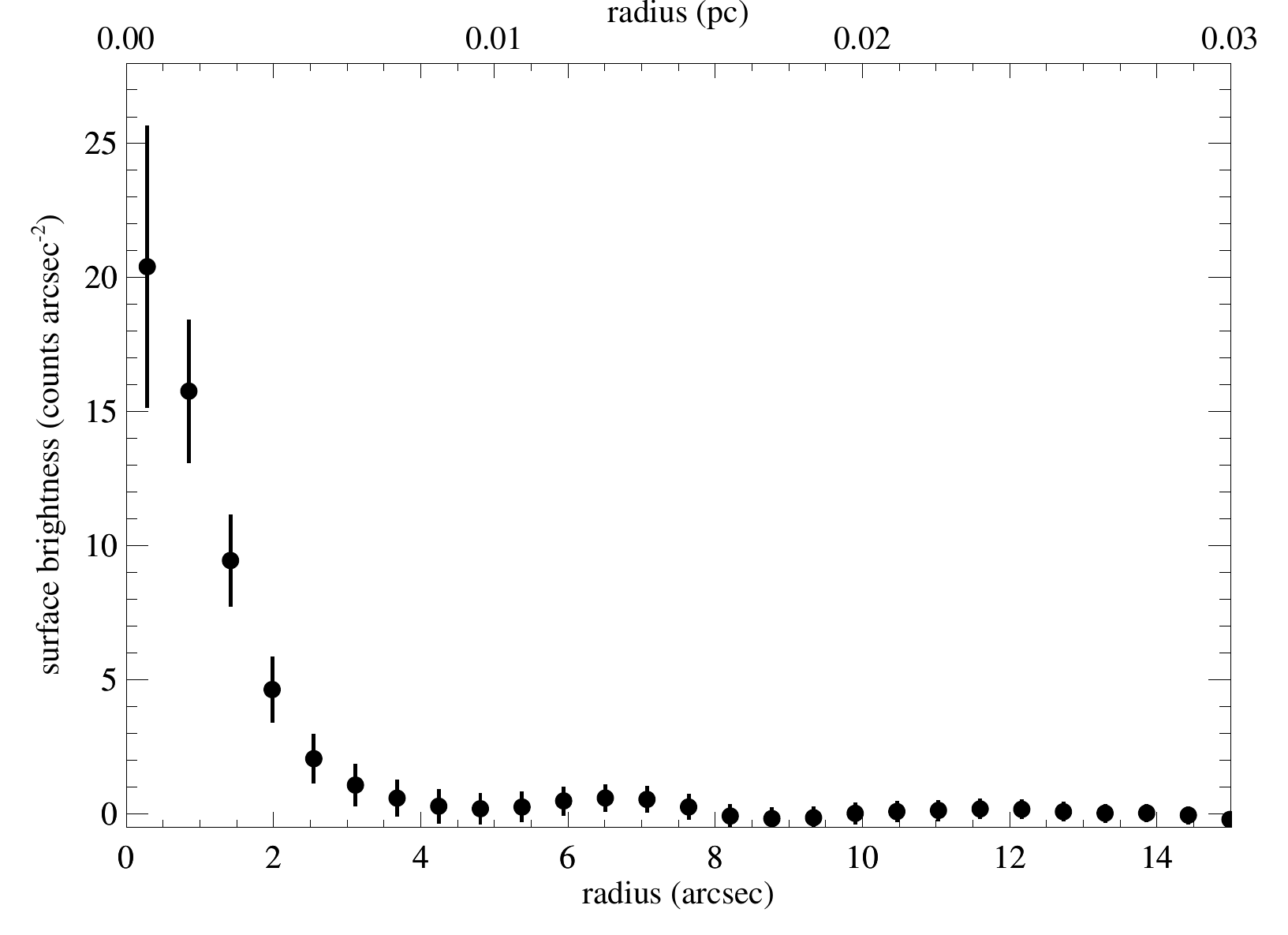}
\end{tabular}
\caption{Simulated 100 ks {\em Lynx} observations (0.2-7 keV) of the HD 34078 (at a distance of 0.41 kpc) stellar wind bow shock \textit{just} below the flux detection limit set by {\em Chandra}. The left panel shows the smoothed ``raw'' counts image. The color bar shows the pixel value, ranging from 0 (black) to $\sim3$ (white). The right panel shows the background-subtracted 0.2-7 keV surface brightness profile extracted from the simulated image, showing the excess X-ray emission is clearly detected. \label{figure:lynx}}
\end{figure*}

Figure~\ref{figure:lynx} shows the results of a 100 ks {\em Lynx} exposure of the HD 34078 bow shock just below our {\em Chandra} detection limit. An X-ray excess clearly evident in the radial surface brightness profile; the bow shock is detectable with a signal-to-noise ratio of $\sim$10. The same bow shock placed farther away ($\sim$2 kpc) is detectable with a signal-to-noise ratio of $\sim$3 in $\sim$500 ks.

\section{Conclusions}\label{section:conclusions}
We have performed a stacking analysis leveraging 2.61 Msec of archival {\em Chandra} observations to search for faint X-ray emission from Galactic stellar wind bow shocks. Our stacked image shows no evidence for excess emission at the expected bow shock location to a flux limit of $\sim4.4\times10^{-16}$ \flux\, which corresponds to a luminosity of $\sim2\times10^{29}$ \lum\ assuming a distance of $\sim$2 kpc (typical of the bow shocks in our sample). This provides the most stringent observational upper limit in the X-ray regime to date.

We assess the plausibility of detecting faint X-ray emission from bow shocks with the proposed future X-ray mission {\em Lynx}. If stellar wind bow shocks are indeed producing X-rays at just below our {\em Chandra} detection limit, the least obscured cases (e.g., the nearest, or runaways at higher Galactic latitudes) should be detectable by {\em Lynx} with modest exposure times.

\acknowledgments
The authors would like to thank H. A. Kobulnicky, L. K. Townsley, and P. S. Broos for useful discussions and feedback on early versions of this manuscript. Support for this work was provided by the National Aeronautics and Space Administration through {\em Chandra} Award Number AR8-19002X issued by the {\em Chandra} X-ray Observatory Center, which is operated by the Smithsonian Astrophysical Observatory for and on behalf of the National Aeronautics Space Administration under contract NAS8-03060.

%

\vspace{5mm}
\facilities{Chandra (ACIS)}


\software{CIAO \citep[v4.10;][]{CIAO},  \href{http://hea-www.cfa.harvard.edu/~jzuhone/soxs/index.html}{SOXS}
          }


\end{document}